\documentstyle[12pt]{article}
\renewcommand{\theequation}{\thesection.\arabic{equation}}

\newlength{\extraspace}
\newlength{\extraspaces}

\newcommand{\be}{\begin{equation}}
\newcommand{\ee}{\end{equation}}

\newcommand{\newsection}[1]{
\vspace{15mm}
\pagebreak[3]
\addtocounter{section}{1}
\setcounter{subsection}{0}
\setcounter{footnote}{0}
\setcounter{equation}{0}
\begin{flushleft}
{\large\bf \thesection. #1}
\end{flushleft}
\nopagebreak
\medskip
\nopagebreak}

\newcommand{\mo}[1]{\sqrt{(#1/L)^{2} - 1 +2M/R}\,}
\newcommand{\mop}[1]{\sqrt{(#1/L)^{2} -1+2M_{+}/R}\,}
\newcommand{\moh}[1]{\sqrt{(#1/\hat{L})^{2} -1+ 2M/\hat{R}}\,}
\newcommand{\moph}[1]{\sqrt{(#1/\hat{L})^{2} - 1 +2M_{+}/\hat{R}}\,}
\newcommand{\smo}[1]{#1/L - \sqrt{(#1/L)^2 - 1 +2M/R\,}}
\newcommand{\smop}[1]{#1/L-\sqrt{(#1/L)^2-1+2M_{+}/R\,}}
\newcommand{\smoh}[1]{#1/\hat{L}-\sqrt{(#1/\hat{L})^{2}-1+2M/\hat{R}\,}}
\newcommand{\smohp}[1]{#1/\hat{L}-\sqrt{(#1/\hat{L})^{2}-1+2M/\hat{R}\,}}
\newcommand{\pip}{\pi_{L}(\rs+\epsilon)}
\newcommand{\pim}{\pi_{L}(\rs-\epsilon)}
\newcommand{\Rp}{R'(\rs+\epsilon)}
\newcommand{\Rm}{R'(\rs-\epsilon)}
\newcommand{\rmin}{r_{\mbox{{\tiny min}}}}
\newcommand{\sq}[1]{\sqrt{|1-#1|}}
\newcommand{\rs}{\hat{r}}
\newcommand{\rsd}{\dot{\hat{r}}}

\begin{document}

\thispagestyle{empty}

\begin{flushright}
{\sc PUPT} 1490, {\sc IASSNS} 94/61\\August 1994
\end{flushright}
\vspace{.3cm}

\begin{center}
{\Large\bf{Self-Interaction Correction to Black Hole Radiance}}\\[20mm]
{\sc Per Kraus}\\[3mm]
{\it Joseph Henry Laboratories\\[2mm]
 Princeton University\\[2mm]
 Princeton, NJ 08544\\[2mm]
  E-mail: perkraus@puhep1.princeton.edu}
 \\[30mm]

{\sc Frank Wilczek}\\[3mm]
{\it Institute for Advanced Study\\[2mm]
     Princeton, NJ 08540\\[2mm]
     E-mail: wilczek@iassns.bitnet}
 \\[30mm]

\end{center}

\nopagebreak

{\sc Abstract}
We consider the modification of the formulas
for black hole radiation,
due to the self-gravitation of the radiation.  This is done
by truncating the coupled particle-hole system to a small
set of modes, that are plausibly the most significant ones,
and quantizing the reduced system.  In this way we find that
the particles no longer move along geodesics, nor is the
action along the rays zero for a massless particle.  The
radiation is no longer thermal, but is corrected in a definite
way that we calculate.  Our methods can be extended in a
straightforward manner to
discuss correlations in the radiation, or between incoming
particles and the radiation.


\newsection{Introduction}

Black hole radiance \cite{hawking} was originally derived in an approximation
where  the background geometry was given, by calculating the
response of quantum fields to this (collapse) geometry.  In this
approximation the radiation is thermal, and much has been made both
of the supposed depth of this result and of the paradoxes that
ensue if it is taken literally.  For if the radiation is accurately
thermal there is no connection between what went into the hole and
what comes out, a possibility which is difficult to reconcile with
unitary evolution in quantum theory -- or, more simply, with the
idea that there are equations uniquely connecting the past with the
future.   To address such questions convincingly, one must
go beyond the approximation of treating the geometry as given, and
treat it too  as a quantum variable.  This is not easy, and as
far as we know no concrete correction to the original result
has previously
been derived in spite of much effort over more than twenty years.
Here we shall calculate what is plausibly the leading correction
to the emission rate of single particles in the limit of large
Schwarzschild holes, by a method that can be generalized in several
directions, as we shall outline.

There is a semi-trivial fact about the classic results
for  black hole radiation,
that clearly prevents the radiation from being accurately
thermal.  This is the effect that
the temperature
of the hole depends upon its mass, so that in calculating the
``thermal'' emission rate one must know what mass of the black hole
to use -- but the mass is different, before and after the radiation!
(Note that a rigorous identification of the
temperature  of a hot body from its radiation,
can only be made for sufficiently high frequencies,
such that the
gray-body factors approach unity.  But it is just in this limit
that the ambiguity mentioned above is most serious.)  As has
been emphasized elsewhere \cite{psstw}, this problem is particularly
quantitatively
acute for near-extremal holes --- it is a general problem for
bodies with finite heat capacity, and in the near-extremal
limit the heat capacity of the black hole vanishes.

To resolve the above-mentioned
ambiguity, one clearly must allow the geometry to fluctuate,
namely to support black holes of different mass.  Another point
of view is that one must take into account the {\it self-gravitational
interaction\/} of the radiation.

\newsection{Model and Strategy of Calculation}


To obtain a
complete description of a self-gravitating particle it would be necessary to
compute the action for an arbitrary motion of the particle and gravitational
field.  While writing down a formal expression for such an object is
straightforward, it is of little use in solving a concrete problem due to the
large number of degrees of freedom present.  To arrive at a more workable
description of the particle-hole system, we will keep only those degrees of
freedom which are most relevant to the problem of particle emission from
regions of low curvature.  The first important restriction is made by
considering only spherically symmetric field configurations, and treating the
particle as a spherical shell.  This is an interesting case
since black hole
radiation into a scalar field occurs primarily in the s-wave,
and virtual transitions to higher
partial wave configurations are formally suppressed by powers of
$\hbar$\begin{footnote}{Since
we do not address the ultraviolet problems of quantum gravity
these corrections are actually infinite, but one might anticipate that
in gravity theory with satisfactory ultraviolet behavior the virtual
transitions will supply additive corrections of order $\omega^2/\Lambda^2$,
where $\Lambda$ is the effective cutoff, but will not alter the
exponential factors we compute.}\end{footnote}.

Before launching into the detailed calculation, which becomes rather
intricate, it seems
appropriate briefly to describe its underlying logic.
After the truncation to s-wave, the remaining dynamics
describes a
shell of matter interacting with a black hole of fixed mass and
with itself.  (The mass as seen from infinity is the
total mass, including that from the shell variable, and is allowed
to vary.  One could equally well have chosen the
total mass constant, and allowed the hole mass to vary.)
There is effectively
one degree of freedom, corresponding to the position of the
shell, but to isolate it one must choose appropriate
variables and solve
constraints, since the original action
superficially appears to contain much more than
this.  Having done that, one obtains an effective action for the
true degree of freedom.
This effective
action is nonlocal, and its full quantization would require one to
resolve factor-ordering ambiguities, which appears very difficult.
Hence we quantize it semi-classically, essentially by
using the WKB approximation.  After doing this one arrives at a
non-linear first order partial differential equation for the phase of
the wave function.  This differential equation may be solved by the
method of characteristics.  According to this method, one solves
for the characteristics, specifies the
function to be determined along a generic initial surface
(intersecting the
characteristics transversally), and evolves the function away from
the initial surface, by integrating the action along the characteristics.
(For a nice brief account of this,
see \cite{whitham}.)

When the background geometry is regarded as
fixed the characteristics for particle motion are simply the
geodesics in that geometry, and they are essentially
independent of
the particle's
mass or energy --- principle of equivalence --- except that
null geodesics are used for massless particles, and
timelike geodesics for massive particles.  Here we find that the
characteristics depend on the mass and energy in a highly non-trivial
way.  Also the action along the
characteristics, which would be zero for a massless particle
and proportional to the length for a massive particle, is
now a much more
complicated expression.
Nevertheless we can solve the equations, to obtain the
proper modes for our
problem.

Having obtained the modes, the final step is to identify the state
of the quantum field --- that is, the occupation of the modes ---
appropriate to the physical conditions we
wish to describe.  We do this
by demanding that a freely falling observer passing through the
horizon see no singular behavior, and that
positive frequency modes are unoccupied in the distant past.  This,
it has been argued,
is plausibly the appropriate prescription for the state of the
quantum field excited by collapse of matter into a black hole, at
least in so far as it leads to late-time radiation.  Using it, we
obtain a mixture of positive- and negative- frequency modes at
late times, which can be interpreted as a state of radiation from
the hole.  For massless scalar particles, we carry the explicit
calculation
far enough to identify the leading correction to the exponential
dependence of the radiation intensity on frequency.

\newsection{Effective Action}

We now derive the Hamiltonian effective action
for a self-gravitating particle
in the s-wave. The Hamiltonian formulation of spherically symmetric gravity
is known as the BCMN model \cite{bcmn}; our treatment of this model follows
that of \cite{polch}.
First, we would like to explain why the Hamiltonian form of the
action is particularly well suited to our problem.
As explained above, our physical problem really contains just
one degree of freedom, but the original action appears to contain
several.  The reason of course is that
Einstein gravity is a theory with constraints and one
should only include a
subset of the spherically symmetric configurations in the
physical description, namely those satisfying the constraints.
In general, in eliminating constraints
Hamiltonian methods are more flexible than Lagrangian
methods.
This appears to be very much the case for our problem,
as we now discuss.

In terms of the variables appearing in the Lagrangian description,
the constraints have the form
$$
{\cal C}_{L} \left[ \rs, \rsd; g_{\mu \nu}, \dot{g_{\mu \nu}} \right] =0,
$$
where $\rs$ is the shell radius, and $ \,\dot{}\,$ represents $\frac{d}{dt}$.
When applied to the spherically symmetric, source free, solutions, one obtains
the content of Birkhoff's theorem -- the unique solution is the Schwarzschild
geometry with some mass, $M$.  Since this must hold for the regions interior
and exterior to the shell (with a different mass $M$ for each), and since $M$
must be time independent, we see that only those shell trajectories which are
``energy conserving'' are compatible with the constraints.  This feature makes
the transition to the quantum theory rather difficult, as one desires an
expression for the action valid for an arbitrary shell trajectory. This defect
is remedied in the Hamiltonian formulation, where the constraints are
expressed in terms of momenta rather than time derivatives,
$$
{\cal C}_{H} \left[ \rs, p; g_{ij}, \pi_{ij} \right] =0.
$$
At each time, the unique solution is again some slice of the Schwarzschild
geometry, but the constraints no longer prevent $M$ from being time dependent.
Thus, an arbitrary shell trajectory $\rs (t)$, $p(t)$, is perfectly consistent
with the Hamiltonian form of the constraints, making quantization much more
convenient.

The starting point for the Hamiltonian formulation of gravity
is to write the metric in ADM form \cite{adm}:
\be
ds^{2}=-N^{t}(t,r)^{2} dt^{2} + L(t,r)^{2}[dr +N^{r}(t,r)dt]^{2}
+R(t,r)^{2}[d\theta ^{2} + {\sin{\theta}}^{2} d\phi ^{2}]
\ee
In considering the above form, we have restricted ourselves to spherically
symmetric geometries at the outset.  With this choice of variables, the action
for the shell is written
\be
S^{s}=-m\int\! \sqrt{-\hat{g}_{\mu \nu} d\hat{x}^{\mu} d\hat{x}^{\nu}}
=-m\int\! dt\, \sqrt{\hat{N}^{t^{2}} -\hat{L}^{2}\left(\rsd+
\hat{N}^{r}\right)^{2}},
\ee
$m$ representing the rest mass of the shell,
and the carets instructing one to evaluate quantities at the
shell $\left(\hat{g}_{\mu \nu}=g_{\mu \nu}(\hat{t},\rs)\right)$.

The action for the gravity-shell system is then
\be
S= \frac{1}{16\pi} \int\!d^{4} x\, \sqrt{-g}\, {\cal R} - m \int\! dt\,
\sqrt{({\hat{N}}^{t})^{2} - {\hat{L}}^{2}(\rsd + {\hat{N}}^{r})^{2}}
+ \mbox{ boundary terms}
\ee
and can be written in canonical form as
\be
S=\int\! dt\, p\, \rsd\, + \int\! dt\, dr\, [\pi_{R}\dot{R} + \pi_{L}\dot{L}
-N^{t}({\cal H}_{t}^{s}+{\cal H}_{t}^{G}) - N^{r}({\cal H}_{r}^{s}+{\cal H}_{r}
^{G})] - \int\! dt\, M_{\mbox{{\scriptsize ADM}}}
\label{act}
\ee
with
\be
{\cal H}_{t}^{s}=\sqrt{(p/\hat{L})^{2}+m^{2}\,}\: \delta(r-\rs)
\ \ \ \ \mbox{;} \ \ \ \
{\cal H}_{r}^{s}=-p\, \delta(r-\rs)
\ee
\be
{\cal H}_{t}^{G}=\frac{L {\pi_{L}}^{2}}{2R^2} - \frac{\pi_{L} \pi_{R}}{R}
+\left(\frac{RR'}{L}\right)' - \frac{{R'}^{2}}{2L} - \frac{L}{2}
\ \ \ \ \mbox{;} \ \ \ \
{\cal H}_{r}^{G}=R'\pi_{R} - L \pi_{L}'
\ee
where $'$ represents $\frac{d}{dr}$.
$M_{\mbox{{\scriptsize ADM}}}$ is the ADM mass of the system.
The inclusion of this last
term deserves some comment.  Because the Einstein-Hilbert action contains
second derivatives, a general variation of the metric variables gives a nonzero
result even when the equations of motion are satisfied and the variation of
the metric is zero on the boundary of the space.  If we restrict the class of
metrics to those which are asymptotically flat, with $N^{t} \rightarrow 1$
and $N^{r} \rightarrow 0$ as $r \rightarrow \infty$, then the variation of the
term $\int dt M_{\mbox{{\scriptsize ADM}}}$
precisely cancels the unwanted terms, and so
gives a well defined variational principle \cite{regge}.
It is important to note that
$M_{\mbox{{\scriptsize ADM}}}$ is a function
of the metric variables (whose explicit form will
be displayed in due course) and is numerically equal to the total mass of the
combined gravity-shell system.

We now wish to eliminate the gravitational degrees of freedom in order to
obtain an effective action which depends only on the shell variables.  To
accomplish this, we first identify the constraints which are
obtained by varying with respect to $N^{t}$ and $N^{r}$:
\be
{\cal H}_{t}={\cal H}_{t}^{s}+{\cal H}_{t}^{G}=0 \mbox{ \ \ \ ; \ \ \ }
{\cal H}_{r}={\cal H}_{r}^{s}+{\cal H}_{r}^{G}=0.
\label{con}
\ee
By solving these constraints, and inserting the solutions back into (\ref{act})
we can eliminate the dependence on $\pi_{R}$ and $\pi_{L}$.  We first consider
the linear combination of constraints
\be
0=\frac{R'}{L}{\cal H}_{t} + \frac{\pi_{L}}{RL} {\cal H}_{r}= -{\cal M}'
+\frac{\hat{R}'}{\hat{L}}{\cal H}_{t}^{s}+\frac{\hat{\pi}_{L}}{\hat{R}\hat{L}}
{\cal H}_{r}^{s}
\label{mcon}
\ee
where
\be
{\cal M} = \frac{{\pi_{L}}^{2}}{2R} + \frac{R}{2} - \frac{R{R'}^{2}}{2L^{2}}.
\label{mdef}
\ee
Away from the shell the solution of this constraint is simply ${\cal M}=$
constant.  By considering a static slice ($\pi_{L}=\pi_{R} = 0$), we see that
the solution is a static slice of the Schwarzschild geometry with $\cal{M}$ the
corresponding mass parameter.  The presence of the shell causes ${\cal M}$ to
be discontinuous at $\rs$, so we write
$$
{\cal M}=M \ \ \ \ r<\rs
$$
\be
{\cal M}=M_{+} \ \ \ \ r>\rs.
\ee
As there is no matter outside the shell we also have
$M_{\mbox{{\scriptsize ADM}}}=M_{+}$.
Then, using (\ref{mcon}) and (\ref{mdef}) we can solve the constraints to find
$\pi_{L}$ and $\pi_{R}$:
$$
\pi_{L}=R \mo{R'} \mbox{ \ \ \ ; \ \ \ } \pi_{R} = \frac{L}{R'} \pi_{L}'
\ \ \ \ \ \ \ \ \ \ \ r<\rs
$$
\be
\pi_{L}=R\mop{R'} \mbox{ \ \ \ ; \ \ \ } \pi_{R}=\frac{L}{R'} \pi_{L}';
\ \ \ \ \ \ \ \ \ \ \ r>\rs.
\label{mom}
\ee
The relation between $M_{+}$ and $M$ is found by solving the constraints at the
position of the shell.  This is done most easily by choosing coordinates such
that $L$ and $R$ are continuous as one crosses the shell, and $\pi_{R,L}$
are free of singularities there.  Then, integration of the constraints
across the shell yields
$$
\pip - \pim = -p/ \hat{L}
$$
\be
\Rp - \Rm = - \frac{1}{\hat{R}} \sqrt{p^{2} + m^{2} \hat{L}^{2}}
\label{cons}
\ee

Now, when the constraints are satisfied a variation of the action takes the
form
\be
dS= p\, d\rs + \int dr (\pi_{R} \delta\! R + \pi_{L} \delta\!L) - M_{+}\, dt
\label{var}
\ee
where $\pi_{R,L}$ are now understood to be given by (\ref{mom}),
and $M_{+}$ is determined by solving (\ref{cons}).
We wish to integrate the expression (\ref{var}) to find the action for an
arbitrary shell trajectory.  As discussed above, the
geometry inside the shell is taken to be fixed (namely, $M$ is held constant)
while the geometry outside the shell will vary in order to satisfy the
constraints.  It is easiest to integrate the action by initially varying the
geometry away from the shell.  We first consider starting from an arbitrary
geometry and varying $L$ until $\pi_{R}=\pi_{L}=0$, while holding
$\rs, p, R, \hat{L}$ fixed:
\be
\begin{array}{l}
\int\! dS = \int_{\rmin}^{\infty}\!dr \int_{\pi = 0}^{L}\! \delta\!L\, \pi_{L}
\vspace{4mm}
\\
=\int_{\rmin}^{\rs-\epsilon}\! dr \int_{\pi=0}^{L}\! \delta\!L\, R \mo{R'}
+\int_{\rs+\epsilon}^{\infty}\! dr \int_{\pi=0}^{L}\! \delta\!L\, R \mop{R'}
\\
= \int_{\rmin}^{\rs-\epsilon}\! dr \left[ RL \mo{R'} + RR'
\log{\left|\frac{ \smo{R'}}{\sq{2M/R}}\right|}\right]
\vspace{1mm}
\\
\ \ \ + \int_{\rs+\epsilon}^{\infty}\! dr \left[ RL \mop{R'} +
RR' \log{\left|\frac{\smop{R'}}{\sq{2M_{+}/R}}\right|}\right]
\label{varyL}
\end{array}
\ee
where the lower limit of integration, $r_{{\scriptsize min}}$,
properly extends to the
collapsing matter forming the black hole; its precise value will not be
important.  We have discarded the constant arising from the lower limit
of the $L$ integration.  In the next stage we can vary $L$ and $R$,while
 keeping
$\pi_{R,L}=0$, to some set geometry.  Since the momenta vanish, there
is no contribution to the action from this variation.

It remains to consider nonzero variations at the shell.  If an arbitrary
variation of $L$ and $R$ is inserted into the final expression of (\ref{varyL})
one finds
\be
dS=\int_{\rmin}^{\infty}\! dr\, [\pi_{R} \delta\!R + \pi_{L} \delta\!L]
- \left [\frac{\partial S}{\partial \hat{R}'}(\rs+\epsilon) -
\frac{\partial S}{\partial \hat{R}'}(\rs-\epsilon)\right]d\!\hat{R}
+ \frac{\partial S}{\partial M_{+}} d\!M_{+}.
\label{try}
\ee
Since $R'$ is discontinuous at the shell,
$$
 \frac{\partial S}{\partial \hat{R}'}
(\rs+\epsilon) - \frac{\partial S}{\partial \hat{R}'}(\rs-\epsilon)
$$
is nonvanishing and needs to be subtracted in order that the relations
$$
\frac{\delta S}{\delta R} = \pi_{R} \mbox{ \ \ ; \ \ }
\frac{\delta S}{\delta L} = \pi_{L}
$$
will hold.  From (\ref{varyL}), the term to be subtracted is
\pagebreak
$$
-\left[\frac{\partial S}{\partial \hat{R}'}(\rs+\epsilon) -
\frac{\partial S}{\partial \hat{R}'}(\rs-\epsilon)\right] d\!\hat{R}
\vspace{4mm}
$$
$$
=-d\!\hat{R} \hat{R} \log{\left|\frac{\smoh{\Rm}}{\sq{2M/\hat{R}}}\right|}
$$
\be
\ \ \ +
d\!\hat{R} \hat{R} \log{\left|\frac{\smohp{\Rp}}{\sq{2M_{+}/\hat{R}}}\right|}.
\ee
Similarly, arbitrary variations of $L$ and $R$ induce a variation of $M_{+}$
causing the appearance of the final term in (\ref{try}).  Thus we need to
subtract
\be
\frac{\partial S}{\partial M_{+}} d\!M_{+} =
-\int_{\rs+\epsilon}^{\infty}\! dr\, L \frac{\mop{R'}}{1-2M_{+}/R} d\!M_{+}.
\ee
Finally, we  consider variations in  $p, \rs$, and $t$.  $t$ variations
simply give $dS = - M_{+} dt$. We do not need to separately consider variations
of $p$ and $\rs$, since when the constraints are satisfied their variations
are already accounted for in our expression for $S$, as will be shown.

Collecting all of these terms, our final expression for the action reads
$$
S= \int_{\rmin}^{\rs-\epsilon}\!dr\,\left[ RL \mo{R'}
+\log{\left|\frac{\smo{R'}}{\sq{2M/R}}\right|}\right]
\vspace{2mm}
$$
$$
+\int_{\rs+\epsilon}^{\infty}\! dr\, \left[ RL \mop{R'}
+RR' \log{\left|\frac{\smop{R'}}{\sq{2M_{+}/R}}\right|}\right]
\vspace{2mm}
$$
$$
-\int\! dt\, \dot{\hat{R}} \hat{R}\left [\log{\left|
\frac{\smoh{\Rm}}{\sq{2M/\hat{R}}}\right|}\right.
\vspace{2mm}
$$
$$
  +\left.  \log{\left|\frac{\smohp{\Rp}}{\sq{2M_{+}/\hat{R}}}\right|}\right]
\vspace{2mm}
$$
\be
+ \int\! dt \int_{\rs+\epsilon}^{\infty}\! dr\, \frac{L \mop{R'}}{1-2M_{+}/R}
\dot{M_{+}} - \int\! dt\, M_{+}.
\label{mess}
\ee
To show that this is the correct expression we can differentiate it; then
 it can
be seen explicitly that when the constraints are satisfied (\ref{var}) holds.

We now wish to write the action in a more conventional form as the time
integral of a Lagrangian.  As it stands, the action in (\ref{mess})
is given for
an arbitrary choice of $L$ and $R$ consistent with the constraints.  There is,
of course, an enormous amount of redundant information contained in this
description, since many $L$'s and $R$'s are equivalent to each other through a
change of coordinates.  To obtain an action which only depends on the truly
physical variables $p, \rs$ we make a specific choice for $L$ and
$R$, {\it ie.} choose a gauge.  In so doing, we must respect the condition
$$
\Rp - \Rm = -\frac{1}{\hat{R}} \sqrt{p^2 + m^2 \hat{L}^2}
$$
which constrains the form of $R'$ arbitrarily near the shell.  Suppose we
choose $R$ for all $r>\rs$; then $\Rm$ is fixed by the constraint, but we
can still choose $R$ for $r<\rs-\epsilon$, in other words, away from the
shell.  We will let $R_{<}'$ denote the value of $R'$ close to the shell but
far enough away such that $R$ is still freely specifiable.  We employ the
analogous definition for $R_{>}'$, except in this case we are free to choose
$R_{>}' = \Rp$.

In terms of this notation the time derivative of $S$ is
$$
L=\frac{dS}{dt}=\rsd\hat{R}\hat{L}\left[ \moh{R_{<}'} - \moph{R_{>}'}
\right]
$$
$$
 \ \ \ \ \ - \dot{\hat{R}} \hat{R} \log{\left| \frac{\smoh{\Rm}}{\smoh{R_{<}'}}
\right|}
$$
\be
 \ \ \ \ +\int_{\rmin}^{\rs-\epsilon}\! dr [\pi_{R} \dot{R} +\pi_{L}\dot{L}]
+ \int_{\rs+\epsilon}^{\infty}\! dr [\pi_{R} \dot{R} + \pi_{L} \dot{L}] -M_{+}.
\label{lag}
\ee
At this point we will, for simplicity, specialize to a massless particle
($m=0$) and define $\eta = \pm = \mbox{sgn($p$)}$.  Then the constraints
(\ref{cons}) read
$$
\Rm = \Rp + \frac{\eta p }{\hat{R}}
$$
\be
\moh{\Rm} = \moph{\Rp} + \frac{p}{\hat{L} \hat{R}}.
\label{newcon}
\ee
These relations can be inserted into (\ref{lag}) to yield
$$
L= \rsd \hat{R} \hat{L} \left[ \moh{R_{<}'} - \moph{R_{>}'}\right]
$$
$$
  -\eta \dot{\hat{R}} \hat{R}\log{\left|\frac{R_{>}'/\hat{L}
- \eta \sqrt{(R_{>}'/\hat{L})^{2} -1 +2M_{+}/\hat{R}}}
{R_{<}'/\hat{L} - \eta \sqrt{(R_{<}'/\hat{L})^{2}-1+2M/\hat{R}}}\right|}
$$
\be
 \ \ \ + \int_{\rmin}^{\rs-\epsilon}\! dr [\pi_{R}\dot{R} + \pi_{L}\dot{L}]
+\int_{\rs+\epsilon}^{\infty}\! dr [\pi_{R} \dot{R} +\pi_{L} \dot{L}] - M_{+}.
\label{nlag}
\ee
Now we can use the freedom to choose a gauge to make (\ref{nlag}) appear as
simple as possible.  It is clearly advantageous to choose $L$ and $R$ to be
time independent, so $\pi_{R} \dot{R} + \pi_{L} \dot{L}=0$.  Also, having
$R'=L$ simplifies the expressions further.  Finally, it is crucial that the
metric be free of coordinate singularities.  A gauge which conveniently
accommodates these features is
$$
L=1 \ \ ; \ \ \ R=r
$$
The Schwarzschild geometry in this gauge is reviewed in the Appendix.
It is
considered in more depth in \cite{kw}.

The $L=1, R=r$ gauge reduces the Lagrangian to
\be
L=\rsd [\sqrt{2M \rs} - \sqrt{2M_{+} \rs}] -\eta \rsd \rs
\log{\left|\frac{\sqrt{\rs}-\eta \sqrt{M_{+}}}{\sqrt{\rs}- \eta \sqrt{2M}}
\right|} - M_{+}
\label{laga}
\ee
where $M_{+}$ is now found from the constraints (\ref{newcon}) to be related to
$p$ by
\be
p= \frac{M_{+} - M}{\eta - \sqrt{2M_{+}/r}}.
\ee
The canonical momentum conjugate to $\rs$ obtained from \ref{laga} is
\be
p_{c}= \frac{\partial L}{\partial \rsd} = \sqrt{2M\rs} - \sqrt{2M_{+}\rs}
- \eta \rs \log{\left|\frac{\sqrt{\rs}-\eta \sqrt{2M_{+}}}{\sqrt{\rs}
-\eta \sqrt{2M}}\right|}
\label{pcan}
\ee
in terms of which we write the action in canonical form as
\be
S=\int\! dt [p_{c} \rsd - M_{+}]
\label{Scan}
\ee
which identifies $M_{+}$ as the Hamiltonian.  We should point out that $M_{+}$
is the Hamiltonian only for a restricted set of gauges.  If we look back at
(\ref{nlag}) we see that the terms $\pi_{R} \dot{R} + \pi_{L} \dot{L}$ will in
general contribute to the Hamiltonian.

\newsection{Quantization}

In this section we discuss the quantization of the effective action
 (\ref{Scan}).
First, it is convenient to rewrite the action in a form which explicitly
separates out the contribution from the particle.  We write
$$
M_{+}=M-p_{t}
$$
so
\be
S=\int\! dt [p_{c} \rsd +p_{t}]
\label{Act}
\ee
and the same substitution is understood to be made in (\ref{pcan}).  We have
omitted a
term, $\int\! dt M$, which simply contributes an overall constant to our
formulas.  In order to place our results in perspective, it
is useful to step back and consider the analogous expressions in flat space.
Our results are an extension of
\be
p=\pm\sqrt{{p_{t}}^{2} -m^{2}}
\label{pflat}
\ee
\be
S=\int\! dt [p\dot{r}+p_{t}].
\ee
Indeed, the $G \rightarrow 0$ limit of (\ref{pcan}), (\ref{Scan})
yields precisely these expressions  (with \linebreak $m=0$).
To quantize, one is tempted to
insert the substitutions $p \rightarrow -i\frac{\partial}{\partial r}$,
$p_{t} \rightarrow -i\frac{\partial}{\partial t}$ into (\ref{pflat}),
so as to satisfy the canonical commutation relations.
This
results in a rather unwieldy, nonlocal differential equation.  In this
trivial case we know, of course, that the correct description of the particle
is obtained by demanding locality and squaring both sides of (\ref{pflat})
before substituting $p$ and $p_{t}$.  So for this example  it is
straightforward to move from the point particle description
to the field theory
description, {\it i.e}. the Klein-Gordon equation.
Now, returning to (\ref{pcan})
we are again met with the question of how to implement the substitutions
$p \rightarrow -i \partial$.  In this case the difficulty is more severe;
we no longer have locality as a guiding criterion instructing us how to
manipulate (\ref{pcan}) before turning the $p$'s into differential operators.
This is because we expect the effective action (\ref{Act}) to be nonlocal on
physical grounds, as it was obtained by including the gravitational field of
the shell.

There is, however, a class of solutions to the field equations for which this
ambiguity is irrelevant to leading order, and which is sufficient to determine
the late-time radiation from a black hole.  These are the short-wavelength
solutions, which are accurately described by the geometrical optics, or WKB,
approximation.  Writing these solutions as
$$
\phi(t,r)=e^{i S(t,r)},
$$
the condition determining the validity of the WKB approximation is that
$$
|\partial S| \gg |\partial^{2}S|^{1/2}, \ |\partial^{3} S|^{1/3} \ \ldots
$$
and that the geometry is slowly varying compared to $S$.  In this regime,
derivatives acting on $\phi(t,r)$ simply bring down powers of $\partial S$,
so we can make the replacements
$$
p_{c} \rightarrow \frac{\partial S}{\partial r} \mbox{ \ \ \ ; \ \ \ }
p_{t} \rightarrow \frac{\partial S}{\partial t}
$$
and obtain a Hamilton-Jacobi equation for $S$.  Furthermore, it is well known
that the solution of the Hamilton-Jacobi equation is just the classical action.
So, if $\rs (t)$ is a solution of the equations of motion found by
extremizing (\ref{Act}), then
\be
S\left(t,\rs (t)\right)= S\left(0,\rs (0)\right) +
\int_{0}^{t}\! dt \left[p_{c}\left(\rs (t) \right) \rsd (t) + p_{t}  \right]
\label{phase}
\ee
where
\be
p_{c}\left(0,\rs \right) = \frac{\partial S}{\partial r}\left(0,\rs \right).
\label{pinit}
\ee
Since the Lagrangian in (\ref{Act}) has no explicit time dependence, the
Hamiltonian $p_{t}$ is conserved.  Using this fact, it is easy to verify that
the trajectories, $\rs (t)$, which extremize (\ref{Act}) are simply the null
geodesics of the metric
\be
ds^{2}=-dt^{2}+\left(dr+\sqrt{\frac{2M_{+}}{r}} dt\right)^{2}.
\ee
    From (\ref{geo}) the geodesics are:
$$
\mbox{ingoing: \ \ \ } t+\rs (t)+2\sqrt{2M_{+} \rs (t)} + 4M_{+}
\log{[\sqrt{\rs (t)}+\sqrt{2M_{+}}]}
$$
$$
\hspace{20mm} =\rs (0)+2\sqrt{2M_{+}\rs (0)} +4M_{+}\log{[\sqrt{\rs
(0)}+\sqrt{2
 M_{+}}]}
\vspace{4mm}
$$
$$
\mbox{outgoing: \ \ \ } t-\rs (t) -2\sqrt{2M_{+}\rs (t)}-4M_{+}
\log{[\sqrt{\rs (t)}-\sqrt{2M_{+}}]}
$$
\be
\hspace{22mm}=-\rs (0)-2\sqrt{2M_{+} \rs (0)}-4M_{+}\log{[\sqrt{\rs
(0)}-\sqrt{2
 M_{+}}]}.
\label{geodef}
\ee
$M_{+}$, in turn, is determined by the initial condition $S(0,r)$ according to
(\ref{pcan}) and (\ref{pinit}):
$$
\hspace{-7mm}\mbox{ingoing: \ \ \ } \frac{\partial S}{\partial r}(0,\rs (0))=
\sqrt{2M\rs (0)}-\sqrt{2M_{+}\rs (0)}+\rs (0)\log{\left|\frac{\sqrt{\rs (0)}
+\sqrt{2M_{+}}}{\sqrt{\rs (0)}+\sqrt{2M}}\right|}
$$
\be
\mbox{outgoing: \ \ \ } \frac{\partial S}{\partial r}(0,\rs (0))
=\sqrt{2M\rs (0)}-\sqrt{2M_{+}\rs (0)}-\rs (0)\log{\left|\frac{\sqrt{\rs (0)}
-\sqrt{2M_{+}}}{\sqrt{\rs (0)}-\sqrt{2M}}\right|}.
\label{mpdef}
\ee
Finally, we can use this value of $M_{+}$ to determine $p_{c}(t)$:
$$
\hspace{-7mm}\mbox{ingoing: \ \ \ } p_{c}(t)=\sqrt{2M\rs (t)}-\sqrt{2M_{+}\rs
(t
 )}
+\rs (t)\log{\left|\frac{\sqrt{\rs (t)}+\sqrt{2M_{+}}}{\sqrt{\rs (t)}
+\sqrt{2M}}\right|}
$$
\be
\mbox{outgoing: \ \ \ } p_{c}(t)=\sqrt{2M\rs (t)}-\sqrt{2M_{+}\rs (t)} -\rs (t)
\log{\left|\frac{\sqrt{\rs (t)}-\sqrt{2M_{+}}}{\sqrt{\rs (t)}-\sqrt{2M}}
\right|}.
\label{pcdef}
\ee
These formulas are sufficient to compute $S(t,r)$ given $S(0,r)$.

As will be discussed in the next section, the relevant solutions needed to
describe the state of the field following black hole formation are those with
the initial condition
\be
S(0,r)=kr \ \ \ \ \ \ k>0
\label{sinit}
\ee
near the horizon.  Here, $k$ must be large ($\gg1/M$) if the solution is to be
accurately described by the WKB approximation.  In fact, the relevant $k$'s
needed to calculate the radiation from the hole at late times become
arbitrarily large, due to the ever increasing redshift experienced by the
emitted quanta as they escape to infinity.  We also show in the next section
 that to compute the emission probability of a quantum of frequency $\omega$,
 we are required to find the solution for all times in the region between
$r=2M$ and $r=2(M+\omega)$.  That said, we turn to the calculation of $S(t,r)$
in this region, and with the initial condition (\ref{sinit}).  The solutions
are determined from (\ref{phase}), (\ref{geodef})-(\ref{pcdef}).  Because of
the large redshift, we only need to keep those terms in these relations which
become singular near the horizon.  We then have for the outgoing solutions:
\be
S(t,r)=k\,\rs (0)-\int_{\rs (0)}^{r}\! d\rs\,\rs \log{\left[\frac{\sqrt{\rs}
-\sqrt{2M_{+}}}{\sqrt{\rs}-\sqrt{2M}}\right]} - (M_{+}-M)t
\label{phasecal}
\ee
\be
t-4M_{+}\log{[\sqrt{r}-\sqrt{2M_{+}}]}=
-4M_{+}\log{[\sqrt{\rs(0)}-\sqrt{2M_{+}}]}
\label{tcal}
\ee
\be
k=-\rs (0) \log{\left[\frac{\sqrt{\rs (0)}-\sqrt{2M_{+}}}
{\sqrt{\rs (0)} -\sqrt{2M}}\right]}.
\label{kcal}
\ee

To complete the calculation, we need to invert (\ref{tcal}) and (\ref{kcal})
to find $M_{+}$ and $\rs (0)$  in terms of $t$ and $r$, and then insert these
expressions into (\ref{phasecal}).  One finds that to next to leading order,
$$
\sqrt{2M_{+}}=\sqrt{2M}+(\sqrt{r}-\sqrt{2M})\frac{(e^{k/2M'}-1)e^{-t/4M'}}
{1+(e^{k/2M'}-1)e^{-t/4M"}}
$$
\be
\sqrt{\rs (0)}=\sqrt{2M}+(\sqrt{r}-\sqrt{2M})\frac{e^{(k/2M' -t/4M')}}
{1+(e^{k/2M'}-1)e^{-t/4M'}}
\label{Mr}
\ee
where
\be
M'=M+\sqrt{2M}(\sqrt{r}-\sqrt{2M})\frac{e^{(k/2M-t/4M)}}{1+e^{(k/2M-t/4M)}}.
\label{Mprime}
\ee
Plugging these relations into (\ref{phasecal}) and keeping only those terms
which contribute to the late-time radiation, one finds after some tedious
algebra,
\be
S(t,r)=-(2M^{2}-r^{2}/2)\log{\left[1+e^{(k/2M'-t/4M')}\right]}.
\label{sol}
\ee

\newsection{Results}

We will now discuss the application of these results to the problem of black
hole radiance. We begin by recalling some general features of the quantization
of a scalar field in the presence of a black hole \cite{birrel}.
The quantization
proceeds by expanding the field operator in a complete set of solutions to the
wave equation,
\be
\hat{\phi}(t,r)=\int\! dk \left[ \hat{a}_{k}f_{k}(t,r)+
\hat{a}_{k}^{\dagger}f_{k}^{*}(t,r)\right].
\label{phi}
\ee
There are, however, two inequivalent sets of modes which need to be considered:
those which are natural from the standpoint of an observer making measurements
far from the black hole, and those which are natural from the standpoint of an
observer freely falling through the horizon subsequent to the collapse of the
infalling matter.  The appropriate modes for the observer at infinity are those
which are positive frequency with respect to the Killing time, $t$.
Writing these modes as
$$
u_{k}(r) e^{-i\omega_{k} t},
$$
$\hat{\phi} (t,r)$ reads
\be
\hat{\phi}(t,r)=\int\! dk \left[ \hat{a}_{k}u_{k}(r) e^{-i\omega_{k} t}
+ \hat{a}_{k}^{\dagger}u_{k}^{*}(r) e^{i \omega_{k} t}\right].
\label{phik}
\ee
These modes are singular at the horizon,
$$
\frac{du_{k}}{dr} \ \rightarrow \infty \mbox{ \ \ as \ \ } r \rightarrow 2M.
$$
Symptomatic of this is that the freely falling observer would measure an
infinite energy-momentum density in the corresponding vacuum state,
$$
 \langle 0_{t} | T_{\mu \nu} | {0}_{t} \rangle \rightarrow \infty
\mbox{ \ \ as \ \ } r \rightarrow 2M
$$
where $\hat{a}_{k} | 0_{t}\rangle =0$. However, we do not expect this to be
the state resulting from collapse, since the freely falling observer is not
expected to encounter any pathologies in crossing the horizon, where the local
geometry is entirely nonsingular for a large black hole.
To describe the state resulting from collapse, it is more appropriate to use
modes which extend smoothly through the horizon, and which are positive
frequency with respect to the freely falling observer.
Denoting a complete set of
such modes by $v_{k}(t,r)$, we write
\be
\hat{\phi}(t,r) \int\! dk \left[\hat{b}_{k}v_{k}(t,r)+
\hat{b}_{k}^{\dagger} v_{k}^{*}(t,r)\right].
\label{phih}
\ee
Then, the state determined by
$$
\hat{b}_{k}|0_{v}\rangle=0
$$
results in a non-singular energy-momentum density at the horizon, and so is a
viable candidate.
The operators $\hat{a}_{k}$ and $\hat{b}_{k}$ are related by the Bogoliubov
coefficients,
\be
\hat{a}_{k}=\int\! dk' \left[\alpha_{kk'}\hat{b}_{k'}+\beta_{kk'}\hat{b}_{k'}
\right]
\label{bog}
\ee
where
$$
\alpha_{kk'}=\frac{1}{2\pi u_{k}(r)}\int_{-\infty}^{\infty}\! dt\,
e^{i\omega_{k} t} v_{k'}(t,r)
$$
\be
\beta_{kk'}=\frac{1}{2\pi u_{k}(r)}\int_{-\infty}^{\infty}\! dt\,
e^{i\omega_{k} t} v_{k'}^{*}(t,r).
\label{bcoef}
\ee
The average number of particles in the mode $u_{k}(r)e^{-i\omega_{k}
t}$ is
\be
N_{k}=\langle0_{v}|\hat{a}_{k}^{\dagger} \hat{a}_{k}|0_{v}\rangle
=\int\! dk'\, |\beta_{kk'}|^{2}.
\label{npart}
\ee
If we treat the black hole as radiating for an infinite amount of time, then
$N_{k}$ will be infinite.  To obtain the {\em rate} of emission we can place
the
hole in a large box and use the density of states $d\omega/2\pi$ for
outgoing particles.  Further, if $|\alpha_{kk'}/\beta_{kk'}|$ is independent
of $k'$, as will be shown to be the case, we can use the completeness relation
\be
\int\! dk' \left[ |\alpha_{kk'}|^{2}-|\beta_{kk'}|^{2}\right]=1
\ee
to obtain for the flux of outgoing particles with frequencies between
$\omega_{k
 }$
and $\omega_{k} + d\omega_{k}$,
\be
F(\omega_{k})=\frac{d\omega_{k}}{2\pi}\frac{1}{|\alpha_{kk'}/
\beta_{kk'}|^{2}-1}.
\ee
This gives the flux of outgoing particles near the horizon.  As the particles
travel outwards, some fraction, $1-\Gamma(\omega)$, of them will be reflected
back into the hole by the spacetime curvature.  Thus, for the flux seen at
infinity we write
\be
F_{\infty}(\omega_{k})=\frac{d\omega_{k}}{2\pi}\frac{\Gamma(\omega_{k})}
{|\alpha_{kk'}/\beta_{kk'}|^{2}-1}.
\label{flux}
\ee

Next, we consider the issue of determining the modes $v_{k}(t,r)$.  As stated
above, we require these modes to be nonsingular at the horizon.  Since the
metric near the horizon is a smooth function of $t$ and $r$, a set of such
modes can be defined by taking their behaviour on a constant time surface,
say $t=0$, to be
$$
v_{k}(0,r)\approx e^{ikr} \mbox{ \ \ as \ \ } r\rightarrow 2M.
$$
This is, of course, the initial condition given in (\ref{sinit}).  Now, the
integrals in (\ref{bcoef}) determining the Bogoliubov coefficients depend on
the values of $v_{k}(t,r)$ at constant $r$.  Since $v_{k}$ is evaluated in the
WKB approximation, the highest accuracy will be obtained when $r$  is as close
to the horizon as possible, since that is where $v_{k}$'s wavelength is short.
On the other hand, in calculating the emission of a particle of energy
$\omega_{k}$, we cannot take $r$ to be less than $2(M+\omega_{k})$, since the
solution $u_{k}(r)e^{-i\omega_{k} t}$ cannot be extended past that point.
Therefore, we calculate the integrals with $r=2(M+\omega_{k})$.

The results of the previous section give us an explicit expression
for $v_{k}$. From (\ref{sol}),
\be
v_{k}(t,2(M+\omega_{k}))=
e^{iS(t,2(M+\omega_{k}))}=
e^{i(4M\omega_{k}+2\omega^{2})\log{[1+e^{(k/2M'-t/4M')}] }}
\ee
where $M'$ is
\be
M'=M+\sqrt{2M}(\sqrt{2(M+\omega_{k})}-\sqrt{2M})\frac{e^{(k/2M-t/4M)}}
{1+e^{(k/2M-t/4M)}}\approx M+\omega_{k}
\frac{e^{(k/2M-t/4M)}}{1+e^{(k/2M-t/4M)}}.
\ee
Then, the integrals are,
\be
\int_{-\infty}^{\infty} dt e^{i\omega_{k} t}e^{\pm
i(4M\omega_{k}+2\omega_{k}^{2
 })
\log{[1+e^{(k/2M'-t/4M')}]}},
\label{integral}
\ee
the upper sign corresponding to $\alpha_{kk'}$, and the lower to $\beta_{kk'}$.
 We
can compute the integrals using the saddle point approximation.
It is readily seen
that for the upper sign, the saddle point is reached when
$$
e^{(k/2M'-t/4M')}
\rightarrow \infty,
$$
 so $t$ is on the real axis.  For the lower sign, the saddle
point is
$$
e^{(k/2M'-t/4M')} \approx -1/2,
$$
which, to zeroth order in $\omega_{k}$, gives
$$
t=4i\pi M + \mbox{real}
$$
and to first order in $\omega_{k}$, gives
$$
t=4i\pi(M-\omega_{k})+ \mbox{real}.
$$
Inserting these values of the saddle point into the integrands gives for the
Bogoliubov coefficients,
\be
\left|\frac{\alpha_{kk'}}{\beta_{kk'}}\right|
= e^{4\pi(M-\omega_{k})\omega
 _{k}}.
\ee
The flux of radiation from the black hole is given by (\ref{flux}),
\be
F_{\infty}(\omega_{k})=\frac{d\omega_{k}}{2\pi}\frac{\Gamma(\omega_{k})}
{e^{8\pi(M-\omega_{k})\omega_{k}}-1}.
\ee

There is an alternative way of viewing the saddle point calculation, which
provides additional insight into the physical origin of the radiation.  Let us
rewrite the integral (\ref{integral}) as
\be
\int_{-\infty}^{\infty}\! dt e^{i\omega_{k} t \pm i S(t,2(M+\omega_{k}))}.
\label{newint}
\ee
The saddle point is given by that value of $t$ for which the derivative of the
expression in the exponent vanishes:
$$
\omega_{k} \pm \frac{\partial S}{\partial t} (t,2(M+\omega_{k}))=0.
$$
But $\partial S/\partial t$ is just the negative of the Hamiltonian,
$$
\frac{\partial S}{\partial t}=p_{t}=M-M_{+}
$$
so the saddle point equation becomes
$$
M_{+}=M\pm \omega_{k}.
$$
To find the corresponding values of $t$, we insert this relation into
(\ref{tcal}) and (\ref{kcal}):
\be
t=4(M\pm \omega_{k})\log{\left[\frac{\sqrt{2(M+\omega_{k})+\epsilon\,}
-\sqrt{2(M+\omega_{k})\,}}
{\sqrt{\rs (0)}
-\sqrt{2(M\pm\omega_{k})}}\right]}
\label{tsad}
\ee
\vspace{2mm}
\be
k=-\rs (0)\log{\left[\frac{\sqrt{\rs (0)}-\sqrt{2(M\pm\omega_{k})}}
{\sqrt{\rs (0)} -\sqrt{2M}}\right]},
\label{ksad}
\ee
where we have written $\rs=2(M+\omega_{k})+\epsilon$ to make explicit that
$\rs$ must lie outside the point where the solutions $u_{k}(r)$ break down.
We desire to solve for $t$ as $k\rightarrow \infty$.  For the upper choice of
sign, we find from (\ref{ksad}) that
$$
\sqrt{\rs (0)}=\sqrt{2(M+\omega_{k})}+\mbox{O}\!\left(e^{-k/2M}\right),
$$
which, from (\ref{tsad}), then shows that the corresponding value of $t$ is
purely real.

For the lower choice of sign we have,
$$
\sqrt{\rs (0)}=\sqrt{2(M+\omega_{k})}-\mbox{O}\!\left(e^{-k/2M}\right).
$$
Continuing $t$ into the upper half plane, we find from (\ref{tsad}) that
$$
t=4i\pi (M-\omega_{k}) + \mbox{ real.}
$$
These results of course agree with our previous findings.

The preceding derivation invites us to interpret
the radiation as being due to
negative energy particles propagating in imaginary time.  The particles
originate from just inside the horizon, and cross to the outside in an
imaginary time interval $4\pi (M-\omega_{k})$.  This, perhaps, helps clarify
the analogy between black hole radiance and pair production in an electric
field, which, in an instanton approach \cite{affleck},
is also calculated by considering
particle trajectories in imaginary time.

\newsection{Comments}

1.  Let us return to the question of thermality. One might have
guessed that the correct exponential suppression factor could be
the Boltzmann factor for nominal temperature corresponding
to the mass of the hole before the radiation, after the radiation,
or somewhere in between.  Thus one might have guessed that the
exponential suppression of the radiance could take the form
$e^{-\omega/T_{\rm before}}$, $e^{-\omega/T_{\rm after}}$, or
something in between.  Our result, to lowest order, corresponds
to the nominal temperature for emission being equal
to  $T_{\rm after}$.

2.  Our methods clearly generalize in a completely
straightforward manner to other forms of black
holes ({\it e. g}. Reissner-Nordstrom or  dilaton holes)
and to charged or massive scalar fields.
We have undertaken extensive calculations along these
lines, whose results will be reported elsewhere.  It should also
be possible to consider emission into higher partial waves,
though this involves some new issues in treating the collective
modes for rotations of a non-symmetric hole, that we have not yet
investigated seriously.  Similarly, it would be interesting to
consider emission with recoil.

3.  One can also consider the case of two mutually gravitating
particles.  This involves geometries with two shell
discontinuities.  By analyzing this problem, we expect to be
able to
address the
question whether there are non-trivial
correlations in the radiation, or between incoming
particles and subsequent radiation.
There are however claims \cite{thooft,verlinde}
that semiclassical methods become
internally inconsistent, or at least highly singular,
in the latter case.

\pagebreak
\appendix
\section*{Appendix:
Schwarzschild Geometry with $L=1, R=r$.}
\renewcommand{\theequation}{A.\arabic{equation}}
In this gauge the line element for the Schwarzschild geometry reads
\be
ds^{2}=-dt^{2} +(dr + \sqrt{\frac{2M}{r}} dt)^{2} +
r^{2}(d\theta^{2} + \sin{\theta}
^{2} d\phi^{2}).
\label{metric}
\ee
These coordinates are related to Schwarzschild coordinates,
\be
ds^{2}=-(1-\frac{2M}{r})dt_{s}^{2} + \frac{dr^{2}}{1-\frac{2M}{r}} +
r^{2}(d\theta^{2}
+\sin{\theta}^{2} d\phi^{2})
\ee
by a change of time slicing,
\be
t_{s}=t-2\sqrt{2Mr} - 2M \log{\left[\frac{\sqrt{r}-\sqrt{2M}}{\sqrt{r}+
\sqrt{2M}}\right]}.
\label{tdef}
\ee
In contrast to the surfaces of constant $t_{s}$, the constant $t$ surfaces pass
smoothly through the horizon and extend to the future  singularity
free of coordinate singularities.

In terms of $r$ and $t$, the radially ingoing and outgoing null
geodesics are given by
$$
\hspace{-10mm}\mbox{ingoing: \ \ \ }
t+r-2\sqrt{2Mr}+4M\log[{\sqrt{r}+\sqrt{2M}]
 }
=v = \mbox{ constant}
$$
\be
\mbox{outgoing: \ \ \ } t-r-2\sqrt{2Mr} -4M\log{[\sqrt{r}-\sqrt{2M}]}=u
=\mbox{ constant }.
\label{geo}
\ee

\end{document}